\documentclass[preprint,showpacs,preprintnumbers,amsmath,amssymb]{revtex4-1}

\usepackage{graphicx}
\usepackage{wrapfig}
\usepackage{times}
\usepackage{etoolbox}
\usepackage{fancyhdr}

\usepackage{amsfonts}
\usepackage{amsmath}
\usepackage{bm}
\usepackage{xcolor}
\usepackage{color}

\usepackage[colorlinks=false,bookmarks=false,citecolor=black,linkcolor=black, pdfstartview=FitH,linktocpage=black]{hyperref}

\begin{document}

\title{The Complex Structure of Magnetic Field Discontinuities in the Turbulent Solar Wind}

\author{A. Greco$^{1}$, S. Perri$^{1}$, S. Servidio$^{1}$, E. Yordanova$^{2}$, and P. Veltri$^{1}$}

\affiliation{$^1$Dipartimento di Fisica, Universit\`a della Calabria, 87036 Rende (CS), Italy}
\affiliation{$^2$Swedish Institute of Space Physics, Uppsala, Sweden}


\input epsf

\begin{abstract}
Using high resolution Cluster satellite observations, we show that the turbulent solar wind is 
populated by magnetic discontinuities at different scales, going from proton down to electron scales.
The structure of these layers resembles the Harris equilibrium profile in plasmas. 
Using a multi-dimensional intermittency technique, we show that these structures are connected through the scales.  
Supported by numerical simulations of magnetic reconnection, we show that observations are 
consistent with a scenario where many current layers develop in turbulence, and where the outflow 
of these reconnection events are characterized by complex sub-proton networks of secondary islands, in a self-similar way.
The present work establishes that the picture of ``reconnection in turbulence'' and ``turbulent reconnection'', 
separately invoked as ubiquitous, coexist in space plasmas.
%The pictures of ``reconnection in turbulence'' and ``turbulent reconnection'', separately invoked 
%as ubiquitous in space plasmas, are now confirmed to coexist.
%The picture where ``reconnection in turbulence'' and ``turbulent reconnection'' coexist, commonly 
%invoked as ubiquitous in space plasmas, is now confirmed by observations.
\end{abstract}

\date{\today}

\maketitle
In the past decades, spacecraft observations suggested that plasma turbulence shares many similarities with 
classical hydrodynamics. The power spectrum of magnetic field fluctuations as a function of frequencies $f$ manifests
an inertial range, with a Kolmogorov-like scaling $f^{-5/3}$ (see for example Ref. \cite{Bruno13}). 
More recently, high resolution measurements 
revealed the presence of a secondary inertial sub-range, 
where the spectrum breaks down and 
exhibits a power index steeper than $-5/3$ \cite{Sahraoui09,AlexandrovaEA09}. 
The characteristic scales at which this break-down occurs
are given by the proton  gyro-radius $\rho_p=v_{th,p}/\Omega_p$ (being $v_{th,p}$ the 
proton thermal speed and $\Omega_p$ the proton gyro-frequency) and/or the proton skin 
depth $d_p=c/\omega_{cp}$ (being $c$ the speed of light and $\omega_{cp}$ the 
proton plasma frequency) \cite{Bale05,Sahraoui09,Sahraoui10,Narita11}. 
At these scales the dynamics
could be mediated by kinetic-Alfv\'en fluctuations, whistler-like perturbations and 
coherent structures such as vortexes and current sheets \cite{OsmanEA14}.

The most narrow current sheets and filaments are present at electron scales, 
where turbulent energy eventually dissipates \cite{Perri12,Goldstein15}, 
even if the energy-dissipation mechanisms in a weakly-collisional plasma such as the turbulent solar wind are far from being understood. 
Recent kinetic simulations \cite{KarimabadiEA13,HaynesEA14,WanEA15} clearly demonstrate that dissipation in 
turbulence takes place at filamentary electron scale current sheets. 
Observations, however, are relatively ambiguous with respect to simulations 
because of measurement limitations (1D data spacecraft samplings).
To elaborate a general picture of the processes that govern plasmas in 
the interplanetary medium as well as in laboratory experiments (e.g, \cite{ShaffnerEA15}), 
it is therefore crucial to characterize the smallest scales with both observations and simulations.

One of the best candidate which may explain plasma energy dissipation on kinetic scales is magnetic reconnection.
Magnetic reconnection is the change of topology of the magnetic field, with subsequent conversion of energy into flows, 
heat, and non-thermal effects. Usually, reconnection and turbulence have been studied as separate topics, but more recently 
it has been suggested that these effects might coexist \cite{ServidioEA09,Goldstein15,OsmanEA14}. Namely, reconnection of thin current 
sheets can take place in magnetohydrodynamics (MHD), as well as in plasma-kinetic models \cite{KarimabadiEA13}. On a 
parallel path \cite{MatthaeusLamkin86,LazarianVishniac99}, 
it has been proposed that the process of magnetic reconnection can be very efficient when 
turbulence develops ``inside'' the above thin current sheets. At very high 
Reynolds number, in fact, it is expected that these narrow  
current layers become strongly unstable, generating micro-plasmoids and secondary 
islands in a self-similar way \cite{Lapenta08,SamtaneyEA09}.

Here we investigate the coexistence of reconnection and turbulence in the space plasmas,
inspecting the solar wind from large to small scales. The analysis of  
high resolution magnetic field measurements will be supported by numerical simulations.
The results obtained merge the picture of ``reconnection-in-turbulence'', 
where large scale energy containing structures reconnect producing layers at scales on the order 
of the proton skin depth, with the picture of ``turbulent-reconnection'', where turbulence 
develops inside the outflows of the above current sheets, at scales on the order of (and much 
smaller than) the proton skin depth.

\begin{figure}
\epsfxsize=8cm
\centerline{\epsffile{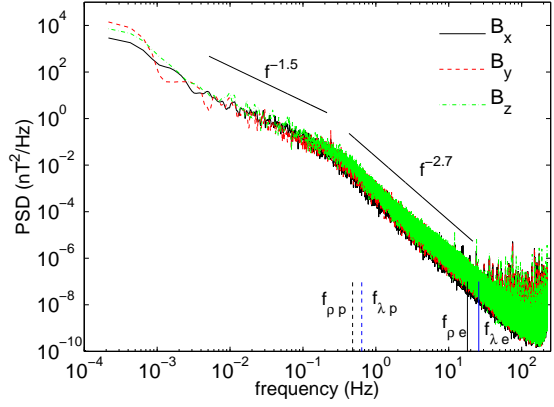}} 
\caption{(Color online) Power spectral density (PSD) of the magnetic field.
The spectral slopes of the inertial range and of the high frequency range are displayed. 
Characteristic plasma frequencies are reported with vertical lines (see text). }
\label{figpwsp}
\end{figure}

The data analyzed are taken in the pristine, undisturbed solar wind from the Cluster 4 
spacecraft on 2007 January 20, 12:00-14:00 UT. Following \citep{Yordanova15}, 
the FGM and the STAFF data in burst mode (sampling frequencies $67$ 
and $450$ vec/sec, respectively) are combined by low-passing FGM data and high-passing 
STAFF data using a cutoff frequency of $1$ Hz. In this sampling, the solar wind bulk speed 
is $V_{\rm sw}\sim 600$km/s, the mean magnetic field (averaged over the entire data set) 
is $B_0\sim 4$ nT (see also Table 1 in \citet{Yordanova15}). 
In terms of characteristic plasma scales, the proton Larmor radius is $\rho_p\sim 193$ km, the
proton inertial length $d_p\sim 163$ km, the electron Larmor radius $\rho_e\sim 3.67$ km and the electron 
inertial length $d_e\sim 3.75$ km. 
The magnetic field power spectral densities of the magnetic field components in Geocentric-Solar-Ecliptic (GSE) 
reference frame are shown in Figure~\ref{figpwsp}. A well defined spectral break 
at $\sim 0.5$Hz is observed, separating the Kolmogorov-like inertial range with spectral slope $-1.5$ from the 
steeper high frequency range. The short vertical lines indicates the Doppler shifted proton 
(electron) gyro-radii $f_{\rho~p(e)}=V_{\rm sw}/2\pi \rho_{p(e)}$ and 
inertial lengths $f_{d~p(e)}=V_{\rm sw}/2\pi d_{p(e)}$. 
The change in the slope of the power spectrum at frequencies 
higher than the proton gyro-frequency indicates 
a change in the nature of the turbulent cascade with possible plasma-kinetic effects at 
work \cite{Sahraoui09,AlexandrovaEA13,ShaffnerEA15}.
However, it is not clear whether the breaking frequency is due to Larmor radius or to proton 
skin depth effects \cite{telloniEA15}.

%%%%%%%%%%%%%%%%%%%%%%%%%%%%%%%%%%%%%%%%%%%%%%%%%%%%%
%    FIGURE 1
%%%%%%%%%%%%%%%%%%%%%%%%%%%%%%%%%%%%%%%%%%%%%%%%%%%%%

\begin{figure}
\epsfxsize=8cm
\centerline{\epsffile{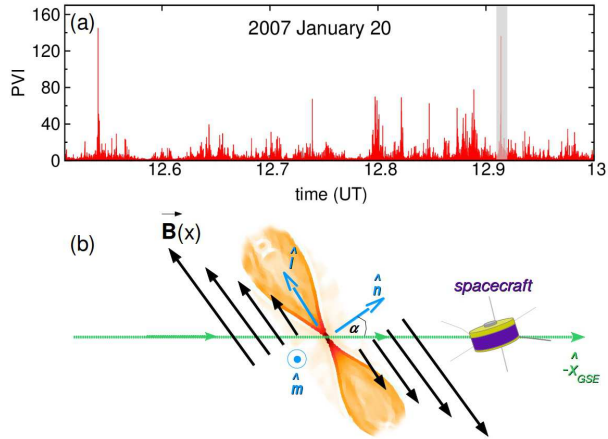}} 
\caption{(Color online) (a) PVI signal computed for $\tau=0.022$ s. (b) Cartoon of a current layer crossing the spacecraft, 
which possibly corresponds to a peak in the PVI. 
The axes of the minimum variance reference system ($\hat{\bf l}, \hat{\bf m}, \hat{\bf n}$) are also depicted.}
\label{figcartu}
\end{figure}

Beside the spectral properties, plasma turbulence is spatially characterized by 
intermittent structures (bursty in space) \cite{Veltri99,BrunoEA01,Kiyani09,Perri12}.
These structures can be classified in several ways \cite{VeltriEA05}, but they are generally strong 
inhomogeneities of the magnetic field \cite{Burlaga69,TS79}. To trace these abrupt spatial changes 
of the magnetic field, we use the Partial Variance of Increments (PVI), which 
measures the ``spikiness'' of the signal relative to a Gaussian value 
and is directly connected to the intensity of the current \cite{DonatoEA13}.
The PVI time series is defined in terms of the magnetic field increment 
vector $\Delta {\bf B}(t, \tau)={\bf B}(t+\tau)-{\bf B}(t)$ \cite{GrecoEA08}:
\begin{equation}
PVI(t,\tau)=\frac{|\Delta {\bf B}(t, \tau)|}{\sqrt \langle |\Delta {\bf B}(t, \tau)|^2 \rangle}
\label{eq:pvi}
\end{equation}
where the average is over a suitably large trailing sample computed along the time series and $\tau$ is the time lag.
For this study we compute PVI on inertial scales from $\tau=15$ s to $1$ s, and on 
kinetic sub-proton scales ranging from $\tau=0.7$ s to $0.022$ s.
The smallest time separation used correspond to a frequency of $45$ Hz, where the signal-to-noise ratio is still high.
The PVI series, computed for $\tau=0.022$ s, is reported in Figure~\ref{figcartu}-(a) for a portion of the data-set. 
The signal, as it can be seen, displays a strongly intermittent character, typical of turbulence.

%%%%%%%%%%%%%%%%%%%%%%%%%%%%%%%%%%%%%%%%%%%%%%%%%%%%%
%    FIGURE 2
%%%%%%%%%%%%%%%%%%%%%%%%%%%%%%%%%%%%%%%%%%%%%%%%%%%%%

In order to characterize the most abrupt events, 
a set of structures with PVI amplitude above a given threshold can be defined. 
Once these structures have been localized, the magnetic filed has been rotated 
in the local minimum variance reference frame \cite{sonnerup,Perri12}, 
defined by the basis $({\hat {\mathbf l}}, {\hat {\mathbf m}, {\hat {\mathbf n}}})$. 
As reported in Figure \ref{figcartu}-(b), $\hat{\mathbf l}$ is the unit vector along the maximum, 
$\hat{\mathbf m}$ along the medium, and $\hat{\mathbf n}$ is along the minimum-variance direction.
Note that in this reference frame the structures can be classified in different ways: tangential 
or rotational discontinuities, magnetic holes and so on \cite{GrecoPerri14} (a detailed 
classification of magnetic discontinuities is beyond the scope of the Letter). 
We further estimate the current density vector ${\bm J}={\bm \nabla}\times{\bm B}/\mu_0$ within those structures. 
From a single-satellite sampling, the only two components of ${\bm J}$ that can be estimated are $J_y$ and $J_z$ 
(multispacecraft techniques are not applicable in this analyzed period).
These components have been computed via the magnetic field differences along GSE $x$ direction, in the limit of small $\tau$. 
Structures showing the minimum variance axis $\hat{\mathbf n}$ almost parallel to $x$ axis have been selected. 
First, we impose that the angle between $\hat{\mathbf n}$ and $x$ is $\alpha<8^{\circ}$. 
Second, we restrict to layers with the minimum variance component $B_{n}\sim B_{m}\sim 0$. 
This configuration restricts the analysis of the current to the cases reported in the cartoon of Figure~\ref{figcartu}-(b).
These constraints ensure that the above surrogate computed from the Cluster 4 data, is very close to the actual current density.

%%%%%%%%%%%%%%%%%%%%%%%%%%%%%%%%%%%%%%%%%%%%%%%%%%%%%
%    FIGURE 3
\begin{figure}
\epsfxsize=8.8cm
\centerline{\epsffile{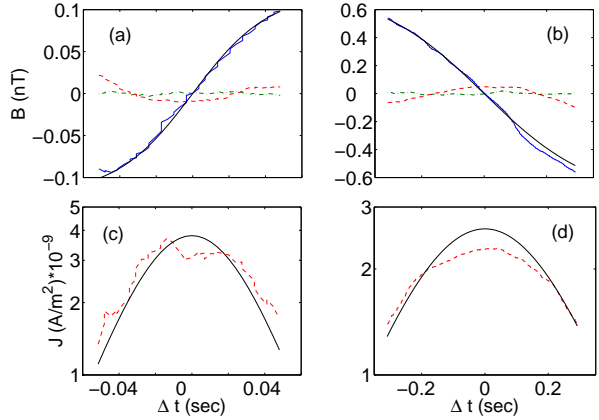}} 
\caption{(Color online) (a) Magnetic field, in the local minimum variance frame, in a discontinuity at electron (a) and sub-proton (b) scales. 
$B_l$ is reported with solid (blue), $B_n$ with dot-dashed (green), and $B_m$ with dashed (red) line.
The thick (black) line represents the Harris profile with $\lambda=3d_e$ (a) and $55d_e$ (b).
(c) and (d) report the surrogate current density (red-dashed) for the discontinuities in (a) and (b), respectively, together with 
the Harris current (black solid).}
\label{figstruct}
\end{figure}
%%%%%%%%%%%%%%%%%%%%%%%%%%%%%%%%%%%%%%%%%%%%%%%%%%%%%

The strongest current layers detected by the PVI method, at time lags that range from inertial to sub-proton time scales
seem to have a self-consistent shape, resembling equilibrium solutions of both fluid-like and kinetic plasmas.
The simplest analytical model that could represent thin current layers on kinetic scales is 
the well-known Harris sheet \cite{Harris},
which is widely used to describe a plasma sheath confined between two regions of oppositely directed magnetic field. 
It is a kinetic equilibrium and the earliest exact analytical solution of the Vlasov equation.
Under some assumptions the magnetic field follows a 1D hyperbolic-tangent profile.
Figure~\ref{figstruct} shows two examples of 1D current sheets, detected by PVI method, 
in their local minimum variance reference frame. 
The time duration is $\Delta t\sim 0.1$ s (a) and $\Delta t\sim 0.7$ s (b), that 
correspond to $60$ km and $420$ km along $x$, respectively, assuming the validity of the Taylor frozen-in hypothesis.
The maximum variance component $B_{l}$ performs a smooth, large amplitude rotation, while $B_{m}\sim B_n\sim 0$. Note that 
$B_{m}$, albeit very small, displays a multi-polar signature, typical of 
solar wind reconnection exhausts \citep{ErikssonEA15}.
The best fits from the Harris model $B=B_0 tanh (x/ \lambda )$, where $x=V_{\rm sw}t$ and $\lambda$ 
is the half-thickness of the layer, are also compared in the same figures, indicating very good agreement with the plasma 
equilibrium theory.
The value of $\lambda $ is $\sim 3 d_e$ for the current sheet on the 
left and $ \sim 55 d_e$ for that on the right. 
In the bottom panels, the magnitude of the (partial) current 
density $J=\sqrt{J_y^2+J_z^2}$ is shown for the same structures, 
comparing the profile with the Harris expectation $J\sim(B_0/\lambda) \cosh^{-2} (x/\lambda)$.

\begin{figure}
\epsfxsize=8cm
\centerline{\epsffile{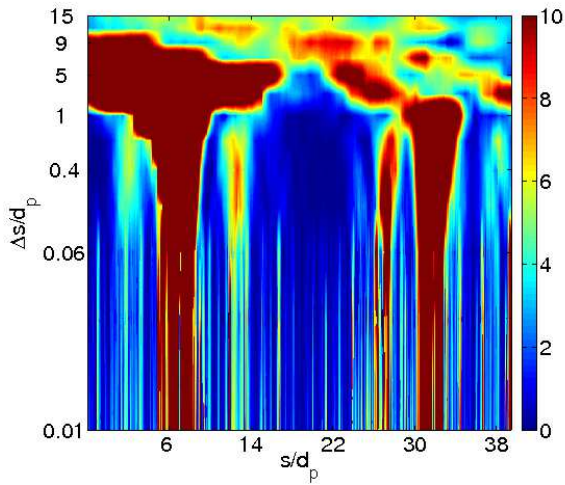}} 
\caption{(Colo online) PVI intensity as a function of space and scale in units of the 
proton inertial length, computed from Cluster 4. Large scale layers break down into 
thinner sub-proton magnetic structures.}
\label{figscalog}
\end{figure}

The Harris profile persists from inertial down to electron scales
in a self similar way. Many of them are isolated, other seem to cluster, 
being embedded in larger scales discontinuous layers. 
In order to quantify this cross-scale connection, we 
performed a multi-dimensional intermittency analysis, computing the 
full PVI as a function of both scales and positions. Figure~\ref{figscalog} represents 
this ``scalogram'' of the PVI series, taken in a sub-interval of the original data-set, 
where time $t$ has been converted in space $s$ and the time lag $\tau $ in the spatial 
scale $\Delta s $, both normalized to the proton inertial length $d_p$. 
The two-dimensional contour clearly shows high values of PVI (current) at large scales 
that connect down to kinetic scales, at wavelengths smaller than the proton skin depth. 
This connection is very interesting and somehow complex, 
being an ubiquitous manifestation of intermittency in plasma turbulence. 
The most interesting feature of the plot is the ``ramification'' 
of the current, following large scale shears, going down to kinetic scales: discontinuities on the order 
of the proton skin depth (or bigger) seem to ``break up'' into smaller sub-proton structures. These nested 
structures are characterized by high values of the current. Some of these are the 1D current sheets depicted 
in Figure~\ref{figstruct}, some other might be related to substructures of the outflows layer, as we will investigate below.

%%%%%%%%%%%%%%%%%%%%%%%%%%%%%%%%%%%%%%%%%%%%%%%%%%%%%
%    FIGURE 4
%%%%%%%%%%%%%%%%%%%%%%%%%%%%%%%%%%%%%%%%%%%%%%%%%%%%%

It has been established, both from observations \cite{RetinoEA07,SundkvistEA07} and from theory and simulations \cite{ServidioEA09},
that turbulence provides a broad distribution of reconnection events, and that at very high 
Reynolds number these micro-structures undergo secondary instabilities, producing turbulent outflows with 
secondary structures embedded \cite{WanEA09}. 
In order to capture the magnetic reconnection at scales
on the order of the proton skin depth, we perform direct numerical simulations 
of two-dimensional (2D) compressible Hall MHD. The use of Hall MHD is important in order to understand the minimal physics of the plasma 
turbulent cascade, as well as the basic physics of reconnection that might be at work in the above intermittent 
events.
We perform the simulation in a periodic geometry, in a $x$-$y$ plane, using for the initial conditions a
double-periodic Harris sheet. The current sheet is modeled such that at the initial time, 
in each current layer, $B_x(y)\sim tanh(y/\lambda )$, with $\lambda\sim 2 d_p$, and sustained by pressure balance.
The code makes use of a spectral algorithm, as described in \cite{GoshEA93}, 
and conserves with high accuracy the global invariants of the system.

%%%%%%%%%%%%%%%%%%%%%%%%%%%%%%%%%%%%%%%%%%%%%%%%%%%%%
%    FIGURE 5
\begin{figure}
\epsfxsize=6.5cm
\centerline{\epsffile{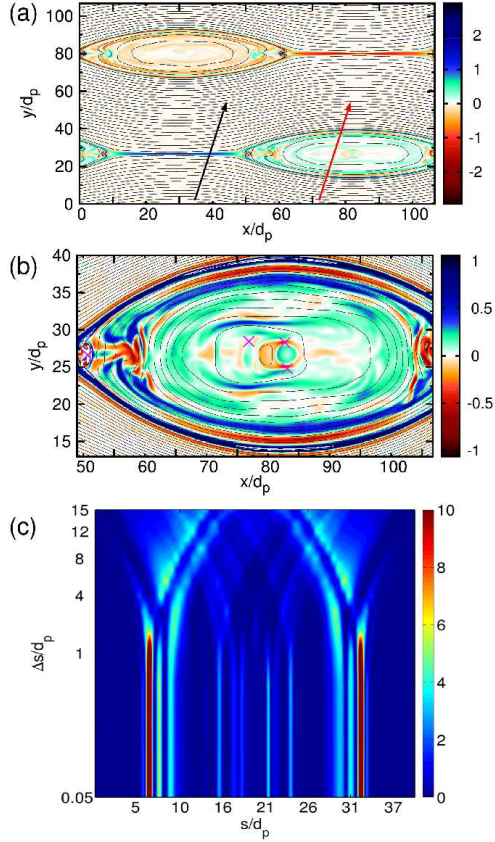}} 
\caption{(Color online) (a) Shaded contour of the current density with in-plane 
magnetic field lines (black). Arrows 
indicate two different possible spacecraft crossings. (b) Zoom into the outflow region, 
showing its complex structure, with secondary 
$\times$-points (magenta symbols). (c) Joint-PVI, as in Figure~\ref{figscalog}, 
for an outflow crossing. The cross-scale coupling at sub-proton scales is evident.
}
\label{figtearing}
\end{figure}
%%%%%%%%%%%%%%%%%%%%%%%%%%%%%%%%%%%%%%%%%%%%%%%%%%%%%

The double-periodic current sheets are perturbed with broad-band noise, with an amplitude of $\sim 5\%$, 
undergoing therefore magnetic reconnection. After the initial evolution, a fully nonlinear steady state of magnetic reconnection 
is reached, as shown in Figure~\ref{figtearing}-(a), where the current density together with the 
in-plane magnetic field lines are reported. Current sheets become thinner
than the proton skin depth, and strong outflows are generated, feeding large scale magnetic islands. 
At this stage of the simulation, we mimic a satellite-like sampling, taking advantage of the periodic boundary conditions.
At the peak of the nonlinear activity, where secondary islands and micro-current
sheets as well as plasmoid have been formed (especially in the exhausts) \cite{LapentaNP15,ErikssonEA15}, 
one can interpolate the magnetic field through a satellite that flows through turbulence, 
simulating the effect of 1D solar wind sampling. These imaginary trajectories are pictorially reported with arrows in 
Figure~\ref{figtearing}-(a). Each trajectory has been chosen to form an oblique angle of $80^\circ $ with $x$, 
cutting the current sheets almost perpendicularly. 
Obviously, even in a very basic and idealized scenario where only current sheets are present, 
there are several possibility for crossing the reconnecting layer, at different angles, and different distances from 
the $\times$-point. Based on simple arguments on the size of the structures 
(small current sheets and big growing islands), the most probable region are the 
outflows \cite{Gosling12}. As reported in Figure~\ref{figtearing}-(b), the structure of the outflows is very complex, 
where different current layers, plasmoids, and secondary current 
sheets are formed. Inside this structure, indeed, one can detect several secondary $\times$-points, 
where reconnection is occurring in a self-similar way.

Finally, in order to  establish a direct connection with observations, we computed $PVI(s,\Delta s)$ on 
an oblique path which crosses the current sheets in the double periodic simulation of Figure~\ref{figtearing}-(a).
As in the solar wind, the imaginary satellite eventually crosses the main current layer, and 
a cross-scale effect in the joint-PVI is observed. In particular, as reported in Figure~\ref{figtearing}-(c), 
when the sampling is along the large outflow, several sub-structures are found in the turbulent layer, down to sub-proton scales.
Obviously, when the scales approach  the electron skin depth, 
other numerical models need to be taken into account, since kinetic-electron physics may play a major role \cite{KarimabadiEA13}.
Current ``fragmentation'' is met along the direction perpendicular to initial current sheets, 
similarly to the scenario observed in solar wind turbulence. Indeed, in plasma turbulence magnetic reconnection
is a common feature, characterizing 
the strongest intermittent current sheets. In these current layers, systematically, 
the sheets break into smaller ones giving rise to a complex network of secondary structures.
These sub-proton structures are here detected with the PVI technique, and the scenario 
is further supported by simulations.

In summary, using Cluster high resolution data, we investigated the structure of thin current sheets that populate the turbulent 
solar wind. The following picture emerges for the solar wind: 
the turbulent cascade naturally forms current sheets at several scales, 
down to the proton skin depth. Approaching smaller scales, a current ``fragmentation'' process
arises, producing Harris-like layers down to scales comparable with the electron skin depth.
These processes have been described with the support of 
2D compressible Hall MHD simulations of magnetic reconnection, where complex outflows are produced, 
showing that the cross-scale structure
of the current sheets, where secondary islands are embedded in the outflows, is consistent with the observations.
The concepts of reconnection-in-turbulence and turbulent-reconnection, that have been invoked 
as separate paradigms of plasma physics in the past decades, are here found to be two synergistic 
processes of space plasmas.

E. Y. received funding from EC FP7 ([7/2007-2013]) under grant agreement No. 313038/STORM.

\end{document}